\documentclass[sigconf]{acmart}
\usepackage{fancybox}

\AtBeginDocument{%
  \providecommand\BibTeX{{%
    \normalfont B\kern-0.5em{\scshape i\kern-0.25em b}\kern-0.8em\TeX}}}

\setcopyright{acmcopyright}
\copyrightyear{2018}
\acmYear{2018}
\acmDOI{XXXXXXX.XXXXXXX}

\acmConference[Conference acronym 'XX]{Make sure to enter the correct
  conference title from your rights confirmation email}{June 03--05,
  2018}{Woodstock, NY}
\acmPrice{15.00}
\acmISBN{978-1-4503-XXXX-X/18/06}

\usepackage{graphicx}
\usepackage{tabularx}
\usepackage{graphicx}
\usepackage{subfig}

\usepackage[ruled,vlined,linesnumbered,noresetcount]{algorithm2e}
\usepackage{threeparttable,color}
\usepackage{booktabs}
\usepackage{multirow}
\usepackage{balance}
\usepackage{bbm,bm}
\usepackage{setspace}
\usepackage{algorithmic}
\usepackage{graphicx}
\usepackage{paralist}

\usepackage{amsmath,amsfonts,enumitem}

\begin{document}

\title{Company Competition Graph}

\author{Yanci Zhang}
\authornote{Disclaimer: Work was done prior to joining Amazon.}
\affiliation{%
  \institution{Wharton Research Data Services}
  \city{Philadelphia}
  \country{USA}
}
\email{yanci@wharton.upenn.edu}

\author{Yutong Lu}
\affiliation{%
  \institution{University of Oxford}
  \city{Oxford}
  \country{UK}
  }
\email{yutong.lu@mansfield.ox.ac.uk}

\author{Haitao Mao}
\affiliation{%
 \institution{Michigan State University}
 \state{Michigan}
 \country{USA}}
\email{haitaoma@msu.edu}

\author{Jiawei Huang}
\affiliation{%
  \institution{Wharton Research Data Services}
  \city{Philadelphia}
  \country{USA}
}
\email{jwhuang@alumni.upenn.edu}

\author{Cien Zhang}
\affiliation{%
  \institution{Wharton Research Data Services}
  \city{Philadelphia}
  \country{USA}
}
\email{cienzhang@alumni.upenn.edu}

\author{Xinyi Li}
\affiliation{%
  \institution{Wharton Research Data Services}
  \city{Philadelphia}
  \country{USA}
}
\email{xinyili@alumni.upenn.edu}

\author{Rui Dai}
\authornote{Corresponding author.}
\affiliation{%
  \institution{Wharton Research Data Services}
  \city{Philadelphia}
  \country{USA}}
\email{rdai@wharton.upenn.edu}

\renewcommand{\shortauthors}{Yanci et al.}

\begin{abstract}
  
 Financial market participants frequently rely on numerous business relationships to make investment decisions. Investors can learn about potential risks and opportunities associated with other connected entities through these corporate connections. Nonetheless, human annotation of a large corpus to extract such relationships is highly time-consuming, not to mention that it requires a considerable amount of industry expertise and professional training. 
 Meanwhile, we have yet to observe means to generate reliable knowledge graphs of corporate relationships due to the lack of impartial and granular data sources. This study proposes a system to process financial reports and construct the public competitor graph to fill the void. Our method can retrieve more than 83\% competition relationship of the S\&P 500 index companies. Based on the output from our system, we construct a knowledge graph with more than 700 nodes and 1200 edges. A demo interactive graph interface is \href{review-graph.ddns.net}{available here}.


\end{abstract}





\maketitle

\section{Introduction} \label{sec:intro}     
Graph is a fundamental data structure that denotes pairwise relationships between entities. 
Graphs are well-utilized in financial research and practice for modeling complex dependency structures in equity markets~\cite{bardoscia2021physics, marti2021review}.
Each node and each edge represent a company and the relationship between companies, respectively.
Previous literature has utilized graphs to represent different relationships between companies, such as edges derived from various data sources \cite{hoberg2016text, mantegna1999hierarchical, tumminello2005tool, gai2010contagion, billio2012econometric, company_as_tribe, lu2022trade, lu2023cotrading}. 
In this paper, we develop a system that automatically extracts corporate graphs of competing companies from the SEC annual reports to ease the usage of deep learning algorithms~\cite{fu2022neuron, mao2021neuron, mao2021source}.


Competition relationship exists when companies operate in the same industry or offer similar products. A stock on one company can diffuse on the competition graph and has an impact on various companies. Therefore, precise and up-to-date competition graphs are crucial for financial practitioners to analyze the cross-impact among companies and make financial decisions. 
For example, the announcement by a leading tech firm to acquire a firm that is a direct competitor of many other tech firms may hurt the stock price of the target's competitors, as investors anticipate increased competition and decreased market share for these companies. 

\begin{figure*}      
  \centering 
  \shadowsize=1mm
    \color{gray}  
    \shadowbox{\fboxsep=0.1mm\fcolorbox{white}{white}{
    \includegraphics[width=0.85\textwidth]{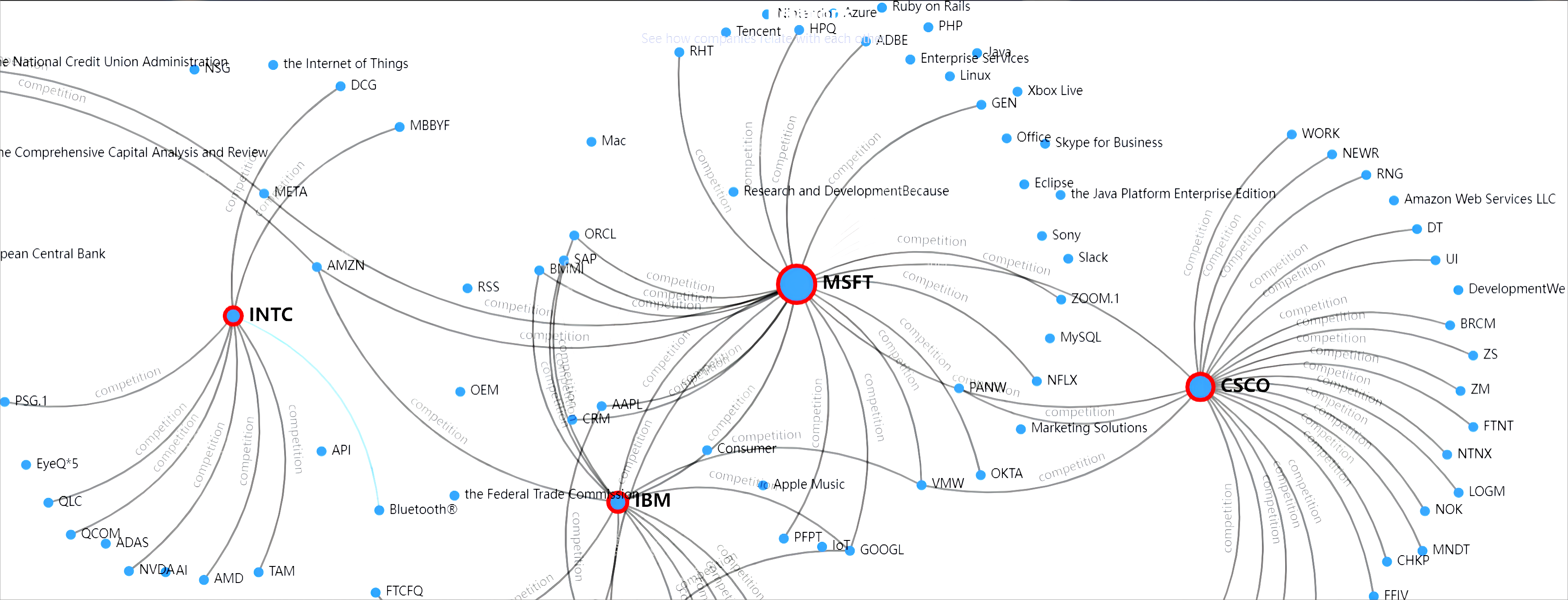}
    }
   }
  \vskip -1em
   
  \caption{
    A snapshot of the competition graph centered around Microsoft. It could be seen the competition relationship between Cisco, Microsoft, IBM and Intel. Solid lines indicate competition relationship. 
  }

  \label{fig:detailed-look}
\end{figure*}

Annual reports of companies are reliable data sources for retrieving competition relations. One of the most popular data sources used to construct competition graph is news, for the reason that it is easy to acquire and relatively short compared to financial filings. However, news as a data source suffers from media attention bias and data quality concerns \cite{fake_news_detection}.
On the contrary, public companies' annual reports are unbiased and granular data sources regulated by the Securities and Exchange Commission (SEC). Many companies explicitly disclose
the names of their competitors in their reports. 
However, the relation extraction from financial reports still heavily relies on professional expertise and industry knowledge. In practice, annual reports are usually long documents with information on companies from all aspects. Moreover, as all US public companies issue annual reports, manual collection of competition relationships among all public companies on a year-by-year basis would be very labor-intensive and time-consuming, if possible.

  
Nonetheless, automating the retrieval of competition relations from annual reports is still under-explored. There are two major challenges. 1) The annual reports of different companies may mention competitors in different places. Furthermore, there may not be any semantic textual information around the names of competing companies. As a result, directly applying name entity recognition (NER) techniques on documents is inappropriate. In absence of semantic information, these techniques cannot differentiate between competitors' names and other name entities, such as the company's products, suppliers, etc., making it challenging to extract competition relations accurately. 2) The reports may use different terminologies or synonyms to refer to the same competitor. Consequently, the constructed competition graphs may contain multiple nodes representing the same companies.

To overcome the aforementioned challenges, we first develop a rule based model to extract subsections that contain only names of competitors from the whole documents. Secondly, we develop customized NLP models, using data labeled by financial experts, to identify name entities representing companies from text and map different names of companies to unique Ids respectively. 
 
Incorporating them as building blocks, we develop an intelligent system to construct competition graphs of companies by automatically extracting competitors directly from their annual reports. Utilizing this system, we build a high-quality competition graph with complete coverage of public companies in the U.S. and accurate edge linkage among companies, with a demonstration interface presented as well.   


\begin{figure}[ht]
  \centering
  \includegraphics[width=\linewidth]{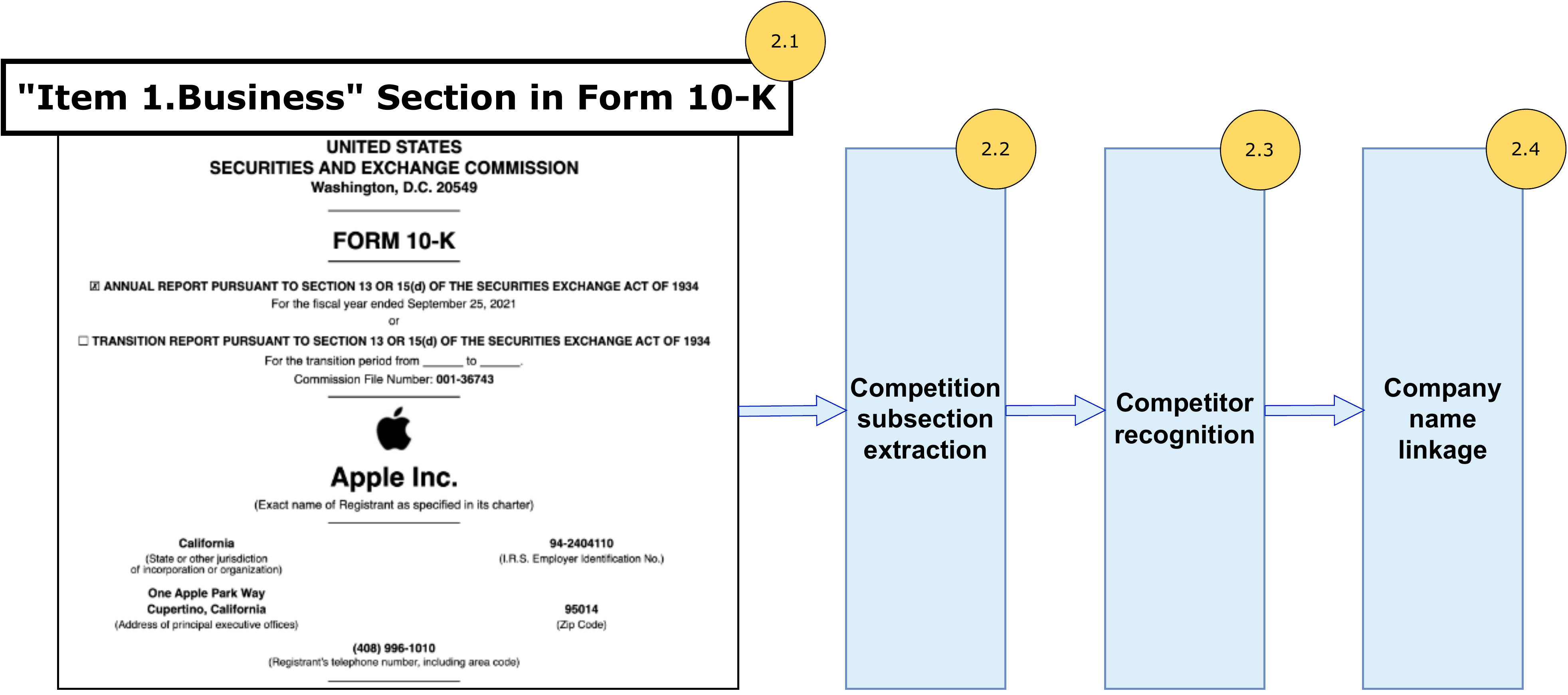}
\vskip -1em
  \caption{Flowchart of the proposed system. The system takes an annual report \cite{aapl2020}, and processes it through three modules in sequential order.
  }
  \label{fig:pipeline}
\end{figure}

\section{System Design} \label{sec:method}
In this section, we introduce the architecture of our system. Figure \ref{fig:pipeline} illustrates an overview of our system. We obtain Form 10-K documents from the EDGAR database (Section \ref{sec:data_description}). The system includes three modules: competition subsection extraction, competing company recognition and competition graph linkage. Given companies' annual reports, the system first locates and retrieves subsections, which disclose competitions that the companies face, from the long documents (Section \ref{sec:subsection_extraction}). Then, the second module recognizes a name entity list of competitors for each company from its competition subsections (Section \ref{sec:company_recognition}). Thirdly, the system accomplishes with a graph by matching company names and then linking competing companies (Section \ref{sec:linkage}). 

\subsection{Input Data} \label{sec:data_description}
  
Annual reports, named Form 10-K, data are acquired from the Electronic Data Gathering, Analysis, and Retrieval (EDGAR) system \footnote{\href{https://www.sec.gov/edgar/searchedgar/companysearch}{https://www.sec.gov/edgar/searchedgar/companysearch}}, 
which is a public database maintained by the Securities and Exchange Commission (SEC). The data are HTML files, which carry format information, such as font, italic, headers and so forth.

Form 10-K discloses financial conditions of the company from all aspects. The document follows hierarchical structures, which usually consist of 22 sections with similar titles, for example "Item 1. Business", prescribed by a guideline from SEC, and then each section includes multiple subsections. In contrast to the consistent structure of sections, companies may organize the subsections in their own ways. The target texts about competition relations concentrate in parts of the Business section with different titles, which we call \textit{competition subsections}. 

\vskip -1em
\subsection{Competition Subsection Extraction} \label{sec:subsection_extraction}
 To handle the aforementioned hierarchical structures, this module extracts the competition subsections in two steps. Firstly, we aim to separate the "Business" section from the rest of the document. To achieve it, we apply a 10-K Itemization System \cite{yancy10q}, which specializes in segmenting the annual report and then reconstructing the sections to exactly follow the SEC guideline.    

After pulling out the Business section, the second step is to apply an original algorithm to accurately locate the starting and ending positions of competition subsections. The algorithm begins with searching for keyword matches, such as "competition", "competitive environment", etc., using regular expressions. With locations of the extracted keywords, the algorithm then determines which are the starting points of the competition sections. The starting point identification relies on format matching, by searching the HTML format information with a set of rules. An example is that the word "competition" appearing in a subtitle with a special font is likely to be a starting point, while not if it is within a sentence. Similarly, the algorithm locates ending points of the competition subsections with a different set of format matching rules.  

\vskip -1em
\subsection{Competitor Recognition} \label{sec:company_recognition}
With retrieved competition subsections, this module recognizes name of competing companies. The competitors’ names can appear in either or both of plain text and table format, in need of distinct ways to handle. Hence, the system detects whether there are "tables" in the text by searching for <tab> in HTML. If so, tables are separated for further processing. There are two main types of tables: bullet points sentences wrapped in <tab> and "real" tables. The system transforms bullet points into plain text by predefined rules. For tables, a table processing algorithm is applied to directly extract company names.  
     
To identify competing companies from plain text, the system applies three name entity recognition (NER) models. 
First, the system calls the NER function from OpenIE \cite{manning2014stanford} by instructing it to extract entity names of type "ORG" which stands for organizations. 
Second, the systems calls Spacy package NER \cite{Honnibal_spaCy_Industrial-strength_Natural_2020} to recognize "COM" entities. 
Third, the system applies a customized BERT-based model \cite{devlin2018bert, liu2019roberta} specialized in distinguishing names of public companies from text. By leveraging Hugging Face \cite{wolf-etal-2020-transformers}, we create a customized model based upon Roberta~\cite{liu2019roberta}. The recognized competitors' names are the union of outputs from the two aforementioned NER models. By taking the union, the system ensures a high recall rate of retrieving correct competitors.

\subsection{Company Name Linkage} \label{sec:linkage} 
The company name linkage module links extracted company names to the publicly traded company identifier. S\&P's Global Company Key~(GVKEY) is selected
\footnote{\href{https://wrds-www.wharton.upenn.edu/pages/grid-items/wrds-sec-linking-tables/}{https://wrds-www.wharton.upenn.edu/pages/grid-items/wrds-sec-linking-tables/}}
as a unique identifier for public traded companies. 
There are two issues with entity names extracted from the previous two stages. Firstly, it is common for companies have multiple names and abbreviations, for example, J.P. Morgan Chase \& Co. can be referred as JPMorgan Chase, JPMorgan, JPM, etc. 
Second, the entity recognition module may extract redundant or incorrect entities names. This module utilizes a pre-trained company name linkage model to resolve them simultaneously, by mapping each of the public company to its GVKEY. Therefore, various names of a company are connected to the same GVKEY.




\section{Experiments} \label{sec:result}
In this section, we implement our system to retrieve the competitors from Form 10-K reports. We structure our experiments and results to answer the following questions. 
\begin{itemize}
    \item How efficient our system is on extracting all the entities?
    \item How does the NER model perform on both the experiment and production scenarios?
    \item How well is the graph constructed through the proposed system?
\end{itemize}

\subsection{Data Preparation} \label{sec:data_preparation}
We collect three datasets for model training in the system: 1) financial reports dataset for Roberta pre-training; 2) publicly traded companies as named entities for Roberta fine-tuning; 3) knowledge base mapping company name to ID. The first two datasets are utilized to train the NLP model with knowledge of publicly traded companies. 
The third one helps the company name linkage model with a knowledge base creating a mapping between the company name and ID. 

For the Roberta pre-training stage, we use textual data from financial reports over the past 30 years. The input data should be in 2 to 5 sentences per data point, with a maximum total token length of 512. The financial report textual data yields a total of 80,928,097 data points. To train the model with more financial context and jargon, this finance domain-specific dataset is fed into the NLP model with Roberta architecture.

For the fine-tuning stage, we create an NER dataset specifically designed for classifying publicly traded companies. 
It contains 68,703 records with over 150,000 entities labeled, original source from WRDS internal databases. 
All the records are labeled by in-house domain experts. The named entities are labeled as positive if they are publicly traded companies. 

For the entity linkage knowledge base, a dataset is created with 114,720 GVKEY and 466,740 corresponding records. The training relies on Spacy linkage training pipeline \cite{Honnibal_spaCy_Industrial-strength_Natural_2020}, where uses total 101,604 sentences for training.

\subsection{Results}
  
We first run experiments on extracting entities from annual reports with and without our proposed system, and report the results in Table \ref{tab:result:inference-speed}. The experiments on inference speed are conducted on a single Nvidia RTX 3090 GPU, with randomly selected filings from Dow 30 index companies. 

We start with applying the NER model from section \ref{sec:subsection_extraction} on the entire annual reports without subsection extraction. On average, it takes 32 minutes to retrieve all the entities from a report. The total computation time for all reports on record would be around 100,000 GPU hours, or 4,200 days.
In comparison, our system takes 5.2 seconds to process a document, which is more than 370 times inference acceleration, thus makes this retrieval task computational feasible to achieve.

\begin{table}[]
\caption{Inference speed comparison between our proposed system and brute force solution.}
\label{tab:result:inference-speed}
\begin{tabular}{c|c|c}
\hline
& Process Whole Filing                                                  & Our system \\ \hline
\#Filing Processes                                                          & 5                                                                     & 5            \\
Total Inference Time                                                        & \begin{tabular}[c]{@{}c@{}}9663 seconds \\ (161 minutes)\end{tabular} & 31 seconds   \\
\begin{tabular}[c]{@{}c@{}}Average Inference Time\\ Per Filing\end{tabular} & 1933 seconds                                                          & 5.2 seconds  \\ \hline
\end{tabular}
\end{table}

We then evaluate the performance of our NER model introduced in section~\ref{sec:company_recognition}. NER performance is crucial to our system's overall performance. If any entities are missing from the output of this step, they will not be included in the graph. Our Roberta-based model has demonstrated solid performance on our initial training dataset. In the actual process of analyzing financial reports, Open IE discovers the largest number of entities. To ensure that we do not miss any entities detected in this step, our final solution combines the results from our Roberta model and the Open IE results, as shown in Table \ref{tab:result:nermodel}.

For labeled training data, the test set consists of 10\% of the data (6,8703 labeled records in total) discussed in section \ref{sec:data_preparation}.
For the results on actual reports, we used the Dow 30 index published between 2018 and 2020, which includes a total of 90 files and represents the actual production data.

\begin{table}[]
\caption{NER model performance on both the training dataset and actual financial reports. In the actual application, each model missed a significant number of named entities. Therefore, we combined the results of our Roberta-based NER model with Open IE to construct the graph further.
}
\label{tab:result:nermodel}
\begin{tabular}{ccccc}
\hline
& \begin{tabular}[c]{@{}c@{}}Our NER \\ Model\end{tabular} & OpenIE & Spacy & \begin{tabular}[c]{@{}c@{}}Our NER \\ + OpenIE\end{tabular} \\ \hline
\multicolumn{5}{c}{Results on Labeled Training Data}                                                                                                                                                                                             \\ \hline
\begin{tabular}[c]{@{}c@{}}Test\\ Recall\end{tabular} & 98\%                                                                      &    80\%    & 79\%  &   -                                                                                      \\ \hline
\multicolumn{5}{c}{Results on Actual reports}                                                                                                                                                                                                    \\ \hline
Precision                                               & 23\%                                                                      & 24\%    & 29\%  & 20\%                                                                                    \\
Recall                                                  & 48\%                                                                      & 71\%    & 47\%  & 72\%                                                                                    \\ \hline
\end{tabular}
\end{table}

We present a summary of the competition graph\footnote{\label{user-interface} User Interface Availble at \url{http://review-graph.ddns.net}} of publicly traded companies, which is the output of our system, in Table \ref{tab:result:graph-summary}. Within S\&P500 annual reports released in 2020, there are 307 companies have competition subsection. Among them, there are 143 companies implicitly mention their competitor names. By linking the annual report publisher and these mentioned competitor names, we have 1295 company to company relationship extracted, which is about 83\% of relationship retrieved.

\begin{table}[]
\caption{Summary of retrieved competition graph. Evaluation statistics is based on the companies from S\&P 500 index. }
\vskip -1em
\label{tab:result:graph-summary}
\begin{tabular}{|c|c|}
\hline
Evaluation Set                                                                                     & S\&P 500       \\ \hline
\#Competition Section                                                                              & 307            \\ \hline
\begin{tabular}[c]{@{}c@{}}\#Filing Mentioned \\  Name of Competitors\end{tabular}                 & 143            \\ \hline
\begin{tabular}[c]{@{}c@{}}\#Competitor\\  Relationship Disclosed\\ (Competitor Edge)\end{tabular} & 1295           \\ \hline
\begin{tabular}[c]{@{}c@{}}\#Competitor Relationship\\ Ground Truth\end{tabular}                   & 1544           \\ \hline
\begin{tabular}[c]{@{}c@{}}\#Public Traded Company\\ (Node)\end{tabular}                           & 685 \\ \hline
\end{tabular}
\end{table}


\begin{figure}[t]
  \centering   
  \shadowsize=1mm
    \color{gray}
    \shadowbox{\fboxsep=0.1mm\fcolorbox{white}{white}{
    \includegraphics[width=0.9\linewidth]{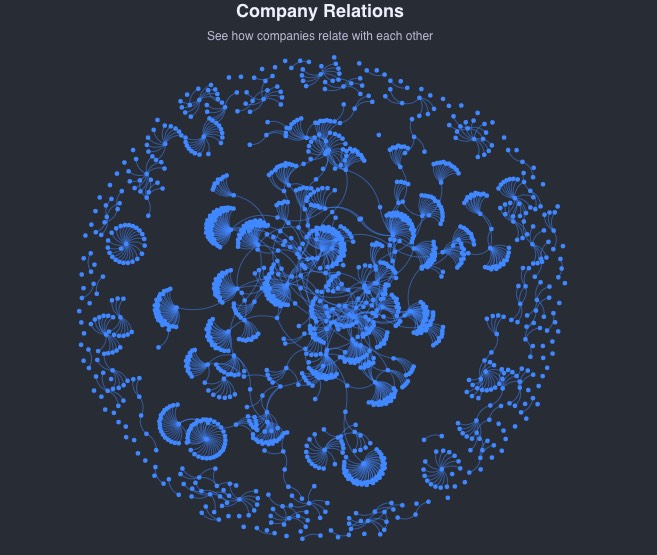}}
    }
    \vskip -1em
  \caption{
    User interface: Competition graph extracted from companies which are constituents of S\&P 500 index.
    }
  \label{fig:graph-overview}
\end{figure}

\section{User Interface} \label{sec:demo}

In this section, a user interface 
is presented to demonstrate our extracted company competition graph.
An overview of the competition relationships among S\&P 500 companies is shown in Figure~\ref{fig:graph-overview}.
It could be observed that some companies are hub nodes, while other companies are clustered around them. 

In addition to the overview across companies, users have an option to select a specific company they are interested in. The selected node would show more details on its connected companies along the path. An illustration can be found in Figure \ref{fig:detailed-look}.

\section{Conclusion \& Future Work} \label{sec:conclusion}
This paper presents a system that outputs the competitive relationships of a company into a knowledge graph. It can help financial researchers and analysts gain valuable insights into the health and performance of individual companies as well as the wider industry. The graph could enable a more complete understanding of the complex network of relationships that exist within the business world, and this knowledge can be used to make better investment decisions.

\bibliographystyle{ACM-Reference-Format}
\bibliography{main}

\end{document}